# Motivating Healthy Water Intake through Prompting, Historical Information, and Implicit Feedback

Davide Neves[1], Donovan Costa[1], Marcio Oliveira[1], Ruben Jardim[1], Ruben Gouveia[1], Evangelos Karapanos[1,2]

[1]Madeira Interactive Technologies Institute, Funchal, Portugal
[2]Cyprus University of Technology, Limassol, Cyprus
`rubahfgouveia@gmail.com, evangelos.karapanos@cut.ac.cy`

**Abstract.** We describe Hydroprompt, a prototype for sensing and motivating healthy water intake in work environments. In a 3-week field deployment of Hydroprompt, we evaluate the effectiveness of three approaches to behavior change: *historical information* enabling users to compare their water intake levels across different times of day and days of week, *implicit feedback* providing subtle cues to users on the current hydration levels, and explicit *prompting* attempting to remind participants when hydration falls below acceptable levels or when substantial amount of time has elapsed since the last sip.

## 1 Introduction

Water is essential to our everyday functioning. Over 50% of the human body consists of water and an average sized person should drink at least 2-3 liters of water per day in order to remain hydrated [1]. However, recent studies have revealed that both adults and children often fail to maintain appropriate levels of hydration throughout the day [3, 4].

One of the most effective triggers for water intake is *thirst*, the sensation of needing to drink [2]. Yet this signal is triggered when there is already a water deficit [2], thus, technology may play a role in establishing healthy water intake habits, and a number of attempts have been made towards this direction. For instance, playful bottle [5] senses water consumption through a mobile phone attached to an everyday drinking mug and attempts to influence users' habits through implicit feedback. Hydracoach (www.hydracoach.com) is a commercial bottle that monitors water consumption and provides historical information such as the average water consumption per hour and the amount of time elapsed since the last sip.

Yet, despite some recent interest on the topic, the domain currently lacks an understanding of the effectiveness of different techniques in motivating healthy water intake, with recent work in the broader area of behavior change tools raising concerns over the long-term effectiveness of existing approaches (see [6,7]).

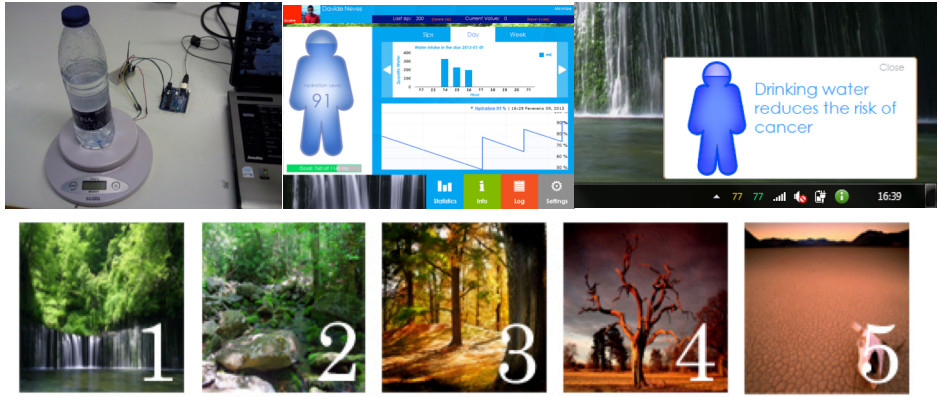

**Figure 1.** The Hydroprompt sensing platform (a) along with the provided feedback: historical information (b), prompting with subliminal feedback (c), implicit feedback (d).

In the current paper we describe the design and evaluation of Hydroprompt, a prototype for sensing and motivating healthy water intake in work environments. In a 3-week field deployment of Hydroprompt, we evaluate the effectiveness of three approaches to behavior change: *historical information* enabling users to compare their water intake levels across different times of day and days of week, *implicit feedback* providing subtle cues to users on the current hydration levels, and explicit *prompting* attempting to remind participants when hydration falls below acceptable levels or when substantial amount of time has elapsed since the last sip. In addition, we attempt to measure the impact of subliminal conditioning [8] (through presenting positive phrases such as "water is good") on the effectiveness of prompting.

## 2  The Hydroprompt System

Hydroprompt employs a load sensor from a regular kitchen scale for sensing water intake (see figure 1). Signal from the load sensor of the scale is led to an operational amplifier and then read by the Arduino platform. This communicates with a C# application that serves as the front-end of Hydroprompt. For successful sensing, the user has to continuously place her bottle or mug on top of the scale. Following the bottle's removal from the scale, the platform senses any difference in weight and records the amount of water intake, if any. The proposed approach offers a simple, cost-effective and reliable sensing platform for water intake applications in stationary settings, such as office environments.

Hydroprompt records the date, time and quantity of each sip and attempts to motivate healthy water intake in three ways: *prompting, historical information* and *implicit feedback*.

**Prompting.** Hydroprompt alerts the user at regular time intervals using a notification at the bottom right part of the user's screen (see figure 1c). The notification consists of a glanceable visualization of the user's hydration level and a short sentence relying on the principle of subliminal conditioning [8]. The idea of subliminal conditioning suggests that priming behavioral concepts (e.g., drinking water) motivates individuals outside conscious awareness, especially when primes match a current need (e.g., fluid deprivation) [9]. A collection of 10 short sentences were created (such as "water is good", "Drinking water helps you feel more energetic", "Drinking water can make you more productive") and were interchanged during different notifications. The frequency of prompting depends on two variables: a) the hydration level of the individual at any given moment, and b) users' stated preference about prompting frequency. Three levels of hydration were defined by the threshold values of 20% and 80%.

**Historical Information.** Hydroprompt presents – using graphs – the amount of water consumed over the course of a week, a day or recent individual sips. In addition, Hydroprompt presents the hydration level of the user at a given moment in time (see figure 3) as well as the extent of goal completion for the day (e.g., 760 of 1140 mL).

**Implicit feedback.** Hydroprompt attempts to provide implicit feedback on individuals' hydration state through altering the wallpaper of their computer environments. The system uses 5 different wallpapers, varying in hydration, to represent five levels of hydration (i.e., above 80%, 60%, 40%, 20%, 0%, see figure 3).

## 3   Field trial of Hydroprompt

We conducted a three-week-long deployment with 6 participants (5 female, all office workers), using an ABA study design, with the goal of assessing the impact of Hydroprompt on individuals' water intake behaviors. The first three and the last three days of the three-week period were used as baseline; during these days, we deployed a stripped-down version of Hydroprompt that sensed water intake without providing any form of feedback to the users. All interactions with the system and water intake behaviors were logged. At the end of each week we conducted an interview with each user to inquire into their experiences and any observed changes in their behavior.

Overall, we observed strong temporal patterns in users' behaviors in all three phases of the study (the two baseline, phase A and phase C, and the intervention, phase B). While Hydroprompt, in Phase B, increased initially the average water intake, this change did not sustain over time, with users quickly reverting to their old behaviors after 6 days. Interestingly, similar temporal trends can also be observed in users the two baseline Phases of the study. These results highlight the role of novelty and increased attention on users' behaviors and stress the importance of longitudinal studies in the establishing the impact of behavior change applications.

To compare the impact of the three intervention approaches (i.e., prompting, historical data and implicit feedback), we contrasted users' behaviors during the five minutes following an event of either three categories (i.e., a notification, an interaction with historical data, and a change in wallpaper).

Users interacted with historical data on a total of 454 times, their desktop wallpaper (implicit feedback) was changed 638 times and received a total of 1383 notifications (prompting). Interacting with historical data was more likely to lead to the user taking a sip of water (with a chance of 30%) than implicit feedback (17%, $\chi^2$=26.9,

p<0.001). Similarly, implicit feedback was more likely to lead to a water intake than prompting (7%, $\chi^2$=46.6, p<0.001).

**Fig. 1.** (left) Water consumption over the course of the study, (right) Number of total and effective events (i.e., events that resulted to the user taking a sip in 5 minutes following the event: a) user's interaction with historical data, b) a change in the wallpaper – implicit feedback, and c) a notification displayed to the user - prompting.

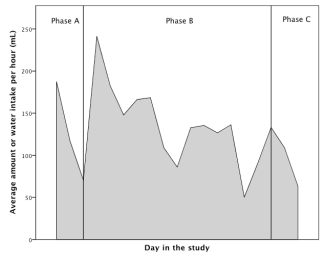

| Type | Effectiveness (Effective out of total events) |
|---|---|
| Historical data | **30,40%** (138 out of 454) |
| Implicit feedback | **17,08%** (109 out of 638) |
| Prompting | **7,16%** (99 out of 1383) |

During our interviews we found users' preferences regarding the different techniques to vary greatly per user. While some participants preferred the notifications and found them not to be intrusive, others commented that they didn't pay attention to them and most participants could not recall the priming messages presented: *"[P4] The notifications get my attention as I'm always busy [thus forgetting to check historical data]", "[P6] I know that the notification shows some messages, but I never notice them"*. Some participants reported the most of their behavior change was induced by implicit feedback, primarily for two reasons. First, wallpaper changes were often visible to the individual's social circle, often leading to discussions around the individual's water consumption. Secondly, some participants reported that the wallpapers impacted their emotions: *"[P2] I feel bad when having 'deteriorating' wallpapers… looking at the screen and seeing the skull, the trees… the other one has a color that forces me to drink"*.